\documentclass[openacc]{rstransa}

\newcommand{\apj}{ApJ}
\newcommand{\aj}{AJ}

\newcommand{\apjl}{ApJ Letters}

\newcommand{\mnras}{MNRAS}
\newcommand{\nat}{Nature}
\newcommand{\icarus}{ICARUS}

\newcommand{\planss}{Planetary Space Science}
\newcommand{\grl}{Geophysics Research Letters}
\newcommand{\jgr}{Journal of Geophysics Research}
\newcommand{\ssr}{Space Science Reviews}

\begin{document}

\title{The Exoplanet Perspective on Future Ice Giant Exploration}

\author{
H.R. Wakeford$^{1}$ and P.A. Dalba$^{2}$}

\address{$^{1}$School of Physics, University of Bristol, HH Wills Laboratory, Tyndall Avenue, Bristol BS8 1TL, UK\\
$^{2}$Department of Earth and Planetary Sciences, University of California Riverside, 900 University Avenue, Riverside CA 92521, USA}

\subject{Extrasolar planets, Space exploration, Solar system}


\corres{Hannah R. Wakeford\\
\email{stellarplanet@gmail.com, hannah.wakeford@bristol.ac.uk}}

\begin{abstract}
Exoplanets number in their thousands, and the number is ever increasing with the advent of new surveys and improved instrumentation. One of the most surprising things we have learnt from these discoveries is not that small-rocky planets in their stars habitable zones are likely common, but that the most typical size of exoplanet is that not seen in our solar system - radii between that of Neptune and the Earth dubbed mini-Neptunes and super-Earths. In fact, a transiting exoplanet is four times as likely to be in this size regime than that of any giant planet in our solar system. Investigations into the atmospheres of giant hydrogen/helium dominated exoplanets has pushed down to Neptune and mini-Neptune sized worlds revealing molecular absorption from water, scattering and opacity from clouds, and measurements of atmospheric abundances. However, unlike measurements of Jupiter, or even Saturn sized worlds, the smaller giants lack a ground truth on what to expect or interpret from their measurements. How did these sized worlds form and evolve and was it different from their larger counterparts? What is their internal composition and how does that impact their atmosphere? What informs the energy budget of these distant worlds? In this we discuss what characteristics we can measure for exoplanets, and why a mission to the ice giants in our solar system is the logical next step for understanding exoplanets.
\end{abstract}


\begin{fmtext}
One big question in science is, \textit{How did we get here?} Planetary Science and Astronomy can help to answer this by placing the Earth and the solar system in which it formed, evolved, and currently lives, into a wider context. To do this we need advanced understanding of our planet and those that occupy the same system to examine the role each evolutionary nuance played in the existence of life able to seek these answers. 
\end{fmtext}

 
\maketitle
\section{Exoplanet demographics and our Solar System}
Exoplanets, planets that orbit stars other than our sun, are now known to exist in their thousands. These worlds are immensely diverse; they do not discriminate between mass, radius, stellar host type, orbital period, and have been discovered in places previously thought of as impossible. 

Characterisation of exoplanets beyond mass and radius is a non-trivial task. To learn about a planet in detail we need to be able to understand the star it orbits, constrain the fundamental parameters of the star-planet system, and measure then interpret the atmosphere of the planet if it has one. 
To do this we need a huge sample of exoplanets to either extrapolate parameters based on models when measurements cannot be made, or the perfect configuration of star, planet, orbit and technology to be able to make the measurements of them directly or through inference. However, this requires a baseline knowledge of what we are looking at, how it is affected by various parameters and what that might mean for the overall understanding or likelihood - is this world a one-off or is it a typical case of formation and evolution?

Out of the thousands of worlds orbiting stars in the Milky way, only $\sim$10\% are likely to be examined in detail. For the coming decades these examinable worlds will be dominated by close-in transiting bodies orbiting bright near-by stars.  
\begin{figure}[!h]
\centering\includegraphics[width=4in]{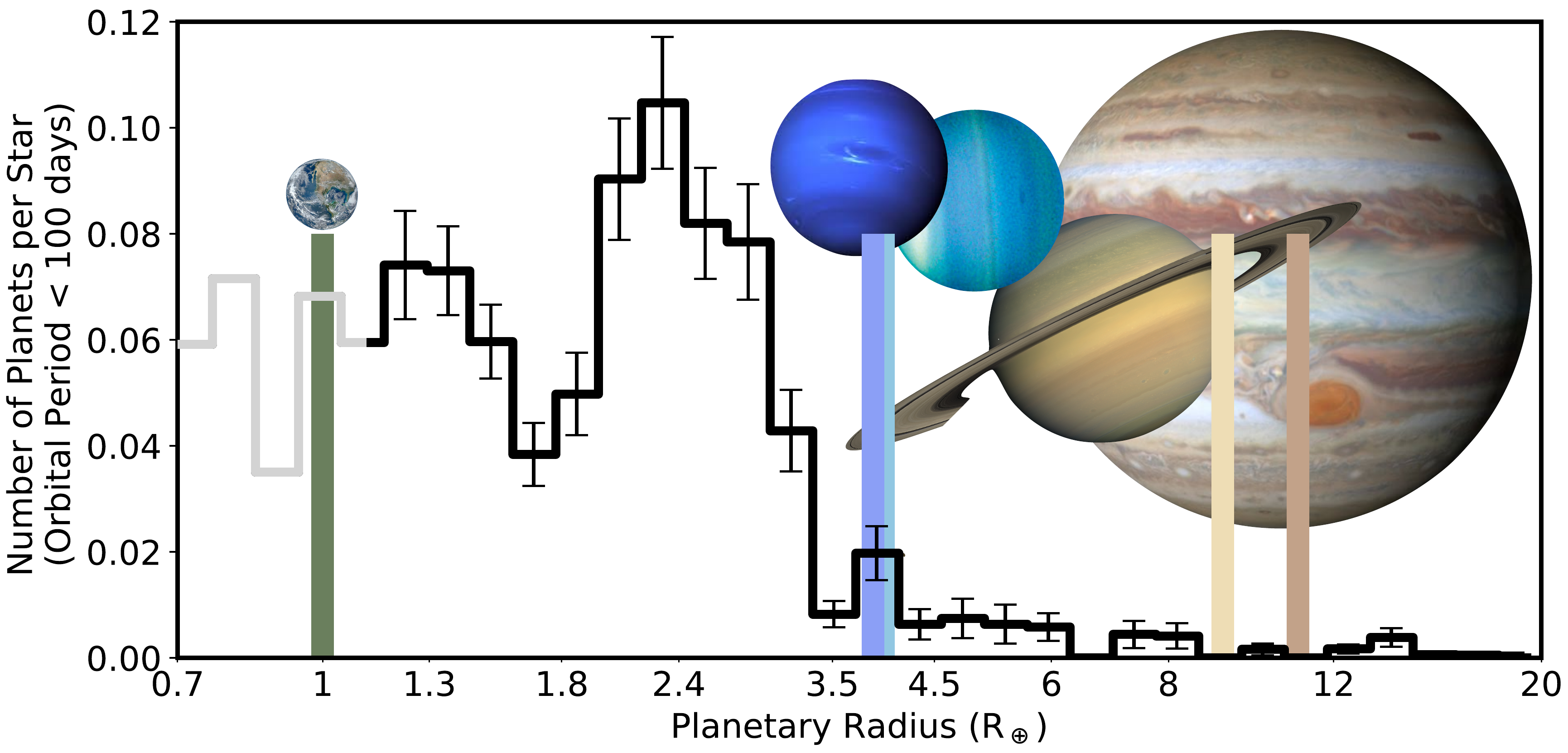}
\caption{The distribution of exoplanet radii discovered on less than 100 day orbits around sun-like stars from the Kepler mission showing the abundance of planets between the size of the Ice Giants and the Earth (remade with permission \cite{fulton2018}).}
\label{fig:fulton}
\end{figure}

If you break the population of exoplanets up into discrete bins according to their radius, the population of exoplanets overwhelmingly sits in a range between that of the Ice Giants, Uranus and Neptune and the terrestrial rocky worlds like the Earth. In Figure\,\ref{fig:fulton} we recreate the radius distribution of exoplanets measured by the \textit{Kepler} Mission \cite{fulton2018}, indicating where in this radius distribution the solar system giants and Earth sit relative to the exoplanet population. While there is a small spike in numbers at around 4\,R$_\oplus$ (4\,Earth radii) the majority of worlds occupy the mini-Neptune (1.8--4\,R$_\oplus$) and super-Earth (1--1.75\,R$\oplus$) regime with a distinct dearth of worlds between. Although, it is important to note that these statistics consider only planets on orbital periods of 100\,days or less orbiting stars with masses between 0.85--1.2M$_\odot$ \cite{fulton2018}.

\begin{figure}[!h]
\centering\includegraphics[width=5.25in]{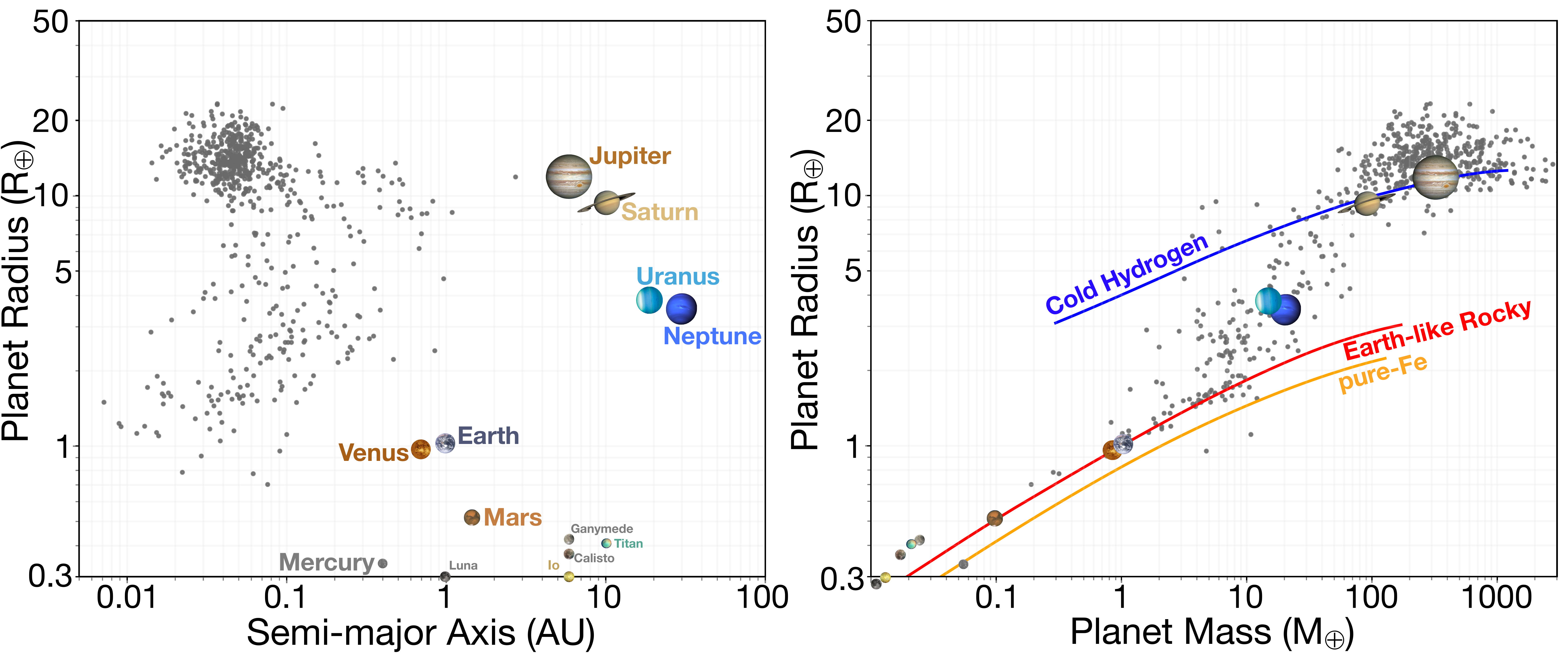}
\caption{Left: Semi-major axis (AU) against planetary radius (Earth radii) for all exoplanets with a measured mass and radius. Right: The same exoplanet data plotted as planetary mass against planetary radius (Earth units). Both show the Solar System planets for comparison. 
Density model lines are shown for cold hydrogen (blue), Earth-like -- 32.5\% Fe + 67.5\% MgSiO$_3$ -- (red), and pure-Fe (orange) \cite{Zeng2019}. This demonstrates the density distribution of planets in our Solar System and their relation to hundreds of measured exoplanets. The Ice Giants lie in a populous region of uncertainty as to their composition and structure, yet are the direct link between the H-dominated gas giants and the terrestrial planets and icy moons of our solar system and exoplanetary systems.}
\label{fig:demographics}
\end{figure}

If we break the population of exoplanets down by considering only the $\sim$10\% of worlds where their mass and radius can be measured, this population reduces (mainly due to technology limits). Even with large scale discovery surveys, such as the microlensing survey planned for WFIRST, still only one of these parameters can normally be determined accurately for any given system. However, the current sample of measurable worlds is still statistically significant to provide needed information on their atmospheres and environment. Figure\,\ref{fig:demographics} shows the population of exoplanets with measured mass and radius as a function of distance from their star and their planetary radius, and in terms of their density distribution from the planetary mass-radius plot. 
These figures highlight some of the observational bias associated with exoplanet detection methods, with detailed atmospheres studies where we have measured the mass and radius of the planet limited to worlds that transit their star from our point of view. However, the statistical distribution of these worlds even on short periods, or around different stars, is a vital factor in placing our solar system worlds into a galactic context.
From these, while it is clear the demographics of exoplanets as a function of distance from their stars does not yet match the solar system planets (or moons), the overall density distribution in many cases can be well explained by looking at the bodies in our solar system. 
There is a clear transition in density from hydrogen-dominated to rock-dominated worlds that is encapsulated by the ice giants of our solar system. 
The best method for filling in the whole phase space to encompass the solar system is through large all sky microlensing surveys, which hold little preference for size, orbital distance, or stellar host, but will be limited to mass only with no opportunity for follow-up. It will therefore be difficult to draw direct lines between the solar system Ice Giants to close-in transiting Neptune-sized worlds.
Yet, we still have little-to-no information about what that ground-truth in bulk density infers for the chemistry, dynamics, and structure of the Ice Giants to draw from.

\section{Ground Truth from Jupiter and Saturn}
We always look to the ``known'' for some ground truth. In planetary science that ``known'' are the planets that make up our solar system. The discoveries made by exoplanets are constantly challenging that ground truth: demographics, formation, composition, etc. Yet, things can only challenge a view on ground truth when there is one to challenge. In the case of the Ice Giants -- Uranus and Neptune -- we do not in fact know the ``knowns''.

The exploration of our solar system has been going on for millennia since the first person looked up and wondered why and how the sun, moon, and stars moved. However, it was not until the invention of the telescope, rockets, and an ability to remotely power objects, that allowed us to truly examine our nearest neighbours. None of which could be done without international collaboration, funding, and a vast number of people with a drive for knowledge.

The \textit{Voyager 2} fly-bys in the late 1980's were the last and only time we have examined our Ice Giants up close, and
our understanding and extent of our solar system is still being realised by the Voyager probes as they leave the influence of our Sun and call another star their home. But these fly-bys only opened questions and answered very little about the most distant planets in our solar system. However, there have been other missions to the other giant planets in our solar system, Jupiter and Saturn, that exoplanet studies are learning from.

In recent years ground truth for our solar system giants has been supplied by the \textit{Cassini} and \textit{Juno} missions to Saturn and Jupiter, respectively. 
\textit{Cassini} was in orbit around Saturn for 13 years from 2004--2017, during that time it got unprecedented views of Saturn, its moons, rings, and atmospheric dynamics. During its grand finale voyage, where it eventually plunged into the atmosphere of Saturn, gravity measurements revealed that unlike Jupiter, Saturn has a deep ($\sim$9000\,km) differential rotation rate with a core mass 15--18\,M$_\oplus$ \cite{Iess2019}.

However, the interior structure of Jupiter may be vastly different than what was theorised only a few years ago. Early results from \textit{Juno}'s gravity science experiment have shown that the core of the planet is not inside a distinct boundary but is extended to a significant fraction of its radius and mixed with a compositional gradient \cite{wahl2017}. 
These new \textit{Juno} measurements prompted another look at the \textit{Cassini} data of Saturn from the Grand Finale mission which may (also!) have a dilute core or compositional gradients \cite{Movshovitz2019}. 
Similarly, a recent re-analysis of the \textit{Voyager 2} fly-bys observations found hints that Neptune, but not Uranus, may be less centrally condensed than previously thought \cite{Nettelmann2013}.
If continued analysis of \textit{Juno} gravity data supports these early results, what is this saying about layers (neatly distributed interiors) in gaseous planets? Will this trend of dilute cores instead of discrete layers continue to Uranus and Neptune? And, what does this mean for their formation and evolution?

Prior to both these missions, the flagship \textit{Galileo} spacecraft and probe explored the atmosphere of Jupiter and the interaction with its moons. The probe itself, that was dropped into the atmosphere in 1995, \textit{seemed} to have fallen into a 'dry' patch detecting very little water during its descent. The Galileo orbiter managed to detect ammonia clouds, and a huge number of lightning strikes caused by the moist convective processes occurring all over the planet \cite{Vasavada2005} perhaps indicating the presence of water vapor, revealing a rich and dynamic atmosphere. 
Subsequently, \textit{Juno} microwave measurements have probed below Jupiter's clouds to reveal that ammonia is not well-mixed below the cloud bases, counter to the expectations of simple thermochemical equilibrium condensation models. Furthermore, \textit{Juno} revealed the presence of a strong ammonia enrichments in the equatorial zone and depletions in the neighboring belts, suggestive of deep circulation patterns that move material from place to place \cite{Bolton2017Sci,Li2017GeoRL}. The vertical and horizontal distributions of water are likely to be similarly surprising.

Both \textit{Cassini} and \textit{Juno} have also revealed that the atmosphere of both worlds are far more dynamic and complex than first thought. Polar orbits have exposed yet more differences in these two-like-planets, with Saturn's pole exhibiting a massive hexagonal polar storm around the planet, while Jupiter's is littered with multiple smaller vortices which mix and collide. This is some-what similar to their apparent equatorial dynamics with Jupiter showing extreme contrast between bands, while Saturn is more uniform in color. These difference in polar to equatorial structure have important implications for future direct imaging exoplanet missions where the orientation and phase of the planet being measured will not be known a priori. 
For the Ice Giants we need to ask the questions: What do the poles/equator of Uranus and Neptune look like, how are they different? What might this imply for future identification of ice giants via direct imaging methods?

\section{Ground Truth Needed from Uranus and Neptune}
To finally complete the detailed picture of our solar system required to truly place our planet and its siblings in context we need to understand the unexplored Ice Giants. To get to the ground truth of these worlds -- their nature and environment both now and in the past -- an in-situ mission is needed to answer a complex series of questions. In the previous sections we have highlighted some of these in relation to Uranus and Neptune, and we are sure other articles in this issue will open many more. Through all of this, however, the overarching question for our solar system and exoplanets remains: 
\emph{\bf How did we get here?}
This relies upon answers to the more detailed questions of: Are our Solar System planets representative of other systems? What role does our Sun play in the planets' energy budget? and How did the Solar System, and exoplanet systems, form and evolve? 

In the following sub-sections we address a number of avenues for advancement in exoplanet studies and gains that can be made by obtaining information on Uranus and Neptune.
The main points we address here are, measuring atmospheric composition, the importance of the interior structure, and formation and evolution markers. 

\subsection{Atmospheric composition and abundances}
The composition and abundance of material in a planetary atmosphere is dictated by a complex interplay of dynamics, radiation (internal and external), and chemistry. Currently, exoplanets with the ability to detect their atmospheres are dominated by those on short period orbits where the planet transits its star from our point of view. 
For transiting exoplanets the amount of information we can obtain about an atmosphere is predominantly governed by the stellar brightness and radius, the planetary radius, temperature (distance from star and type of star), and gravity (mass). Essentially, the cross section of the atmospheric scale height compared to the stellar size and brightness; such that photons passing through the atmosphere can be detected in high enough numbers. However, these factors do not just dictate if a transiting exoplanet atmosphere can be measured, but also what that atmosphere might be composed from and how it is structured. The planetary scale height is an important factor that also relies on the mean molecular weight ($mmw$) of the material in the atmosphere, and therefore the chemistry. This biases observational measurements of exoplanets to those with larger scale heights and therefore planets with lower gravity and higher temperature. However, as the technique and instrumentation have improved investigations have pushed the limits to smaller and longer period planets. While not reaching even remotely near the orbital periods of our solar system, the Neptune-mass "ice giants" with measured atmospheric transmission spectra present an intriguing picture. 

\begin{figure}[!h]
\centering\includegraphics[width=5in]{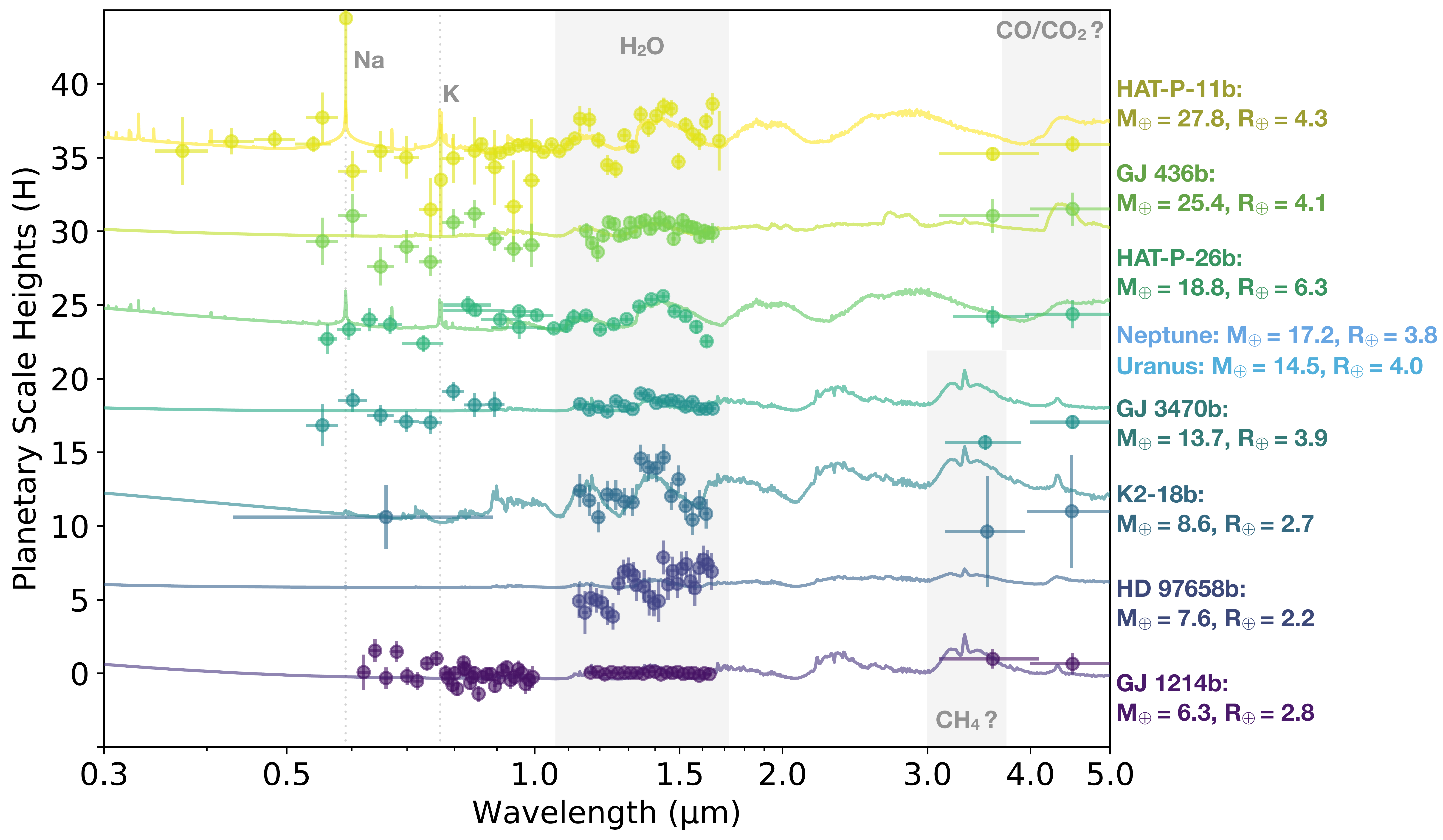}
\caption{Atmospheric transmission spectra for a handful of exoplanets roughly Neptune-sized that have been measured with the Hubble and Spitzer Space Telescopes. Each spectrum has been fit with a model from a grid of forward models scaled to each exoplanets scale height \cite{goyal2019}. Data: HAT-P-11b \cite{chachan2019}, GJ\,436 \cite{knutson2014a}, HAT-P-26b \cite{Wakeford2017Science}, GJ\,3470b \cite{benneke2019}, K2-18b \cite{Benneke2019_K218b}, HD\,97658b \cite{knutson2014c}, GJ\,1214b \cite{bean2011,kreidberg2014a}.}
\label{fig:atmosphere}
\end{figure}

Even though exoplanets are investigated entirely through remote sensing, and often indirectly by means of their star's light, we can get information on their upper atmospheric composition with relative ease compared to our nearest neighboring Ice Giants. Figure\,\ref{fig:atmosphere} shows the transmission/absorption spectrum of seven exoplanets in the Neptune to mini-Neptune mass/radius regime. These planets span a range of temperatures from 250--1000\,K with planetary gravities from 4--14\,ms$^{-1}$. The models used are fully scale-able generic forward models \cite{goyal2019}, to fit these to the data we use a consistent $mmw$ for the atmosphere of 2.7, approximately representing that of Neptune \cite{helled2010interior} but not considered high enough to begin shrinking spectral features \cite{line2016}. While a number of these planets have measured atmospheric metallicities, total abundance of material heavier than hydrogen and helium, indicative of higher or lower $mmw$ \cite{morley2017, Wakeford2017Science,chachan2019}, we keep them all consistent to illustrate trends and features in the exoplanet spectra that can be measured. This series of exoplanets show evidence for absorption due to sodium, potassium, and water vapour in their atmospheres from spectroscopic measurements with the Hubble Space Telescope from the optical to near-IR. Photometric measurements from the Spitzer Space Telescope hint at tentative evidence for CO/CO$_2$ and CH$_4$ in the infrared, however, in many cases, CH$_4$ has not been found when expected \cite{kreidberg2014a,benneke2019}. Almost all of these atmospheres show evidence of aerosols (clouds/hazes) in the atmosphere muting or totally obscuring molecular features. Correlations relating cloudiness to planetary equilibrium temperature have been suggested \cite{crossfield2017} where, as the temperature drops the cloudiness increases. However, new measurements of a distinct water absorption feature in the transmission spectrum of K2-18b (T$_{eq}$ = 265\,K)\cite{Benneke2019_K218b} go against this trend.
Spectroscopic measurements in the IR with the James Webb Space Telescope will greatly expand our understanding of this class of worlds and help push down to smaller and cooler planets more akin to the ice giants and mini-Neptunes. 
Critically the IR spectroscopic capabilities of \textit{Webb} will mean that carbon-based species can definitively be detected in the atmospheres exoplanets over a range of sizes and environments. CO, CO$_2$, and CH$_4$ each have dominant absorption signatures beyond the current capabilities of \textit{Hubble}, with overlapping bands of CO and CO$_2$ around 4--5\,$\mu$m difficult to disentangle at solar abundances in the photometric bandpass of \textit{Spitzer}. \textit{Webb} will additionally be able to measure multiple absorption bands of H$_2$O simultaneously, greatly increasing the constraints that can be placed on its abundance.
These measurements in turn will give us information on the abundance of carbon to oxygen species, the C/O ratio, and upper atmospheric metallicity that can be used to put our own Ice Giants in context with the plethora of similar sized exoplanets in our galaxy. 

Carbon is an important species to characterize in these atmospheres as the C/O ratio is indicative of formation location in the protoplanetary disk \cite{oberg2011} and plays a key role in the atmospheric metallicity determination. 
Ice lines in the planet forming disk, defined by the distance from the star and therefore temperature of the region where different materials such as H$_2$O, CO$_2$, and CO "freeze-out", are thought to be imprinted on the planets forming in those regions via the accumulation and abundance of materials present to form planets. Measurements of the ratios of oxygen-bearing species to carbon-bearing species in the gas phase for these hot giant planets can potentially provide a first order indicator for where the planet originally formed in relation to these ice lines. 
At higher temperatures CO is the dominant carrier of carbon, but at temperatures below 1,000\,K CH$_4$ is expected to become dominant \cite{moses2013}. 
In the Ice Giant atmospheres CH$_4$ is lofted to higher altitudes in Neptune compared to Uranus \cite{moses2017}, which has implications for hydrocarbon chemistry and haze formation. 
These atmospheric processes and haze formation pathways are vital for exoplanet atmospheres, that have already demonstrated cloud/haze properties. 
A study by Gao et al. \cite{Gao2020} shows that for cooler exoplanets, T$_{eq}<$950\,K, hydrocarbon aerosols will be the dominant species present due to an increase in methane abundance when looking at the 1.4\,$\mu$m water vapor absorption feature. The increased presence of photochemically generated species in turn can cause an increase in the opacity of the atmosphere and in the difficulty associated with making measurements of molecular features as aerosols mute and obscure gas phase absorption. 
Additionally, thermal and chemical profiles across various latitudes are useful comparisons against exoplanet retrievals. Many studies have shown that globally averaged thermal profiles are not effective at interpreting transmission \cite{line2016,MacDonald2020} or emission \cite{Feng2016,Taylor2020} spectra. Instead, multiple profiles simulating the integrated 2D/3D environment, including the radiative effects of aerosols, are needed to understand the nature of these exoplanet atmospheres. 

There is substantial precedence for utilizing solar system observations to determine how our home system would appear if viewed from across the Galaxy. Indeed, the transmission spectra of Venus, Earth, Jupiter, Saturn, and Titan have been reconstructed from solar system data sets \cite{Barstow2016,GarciaMunoz2012a,Irwin2014,Dalba2015,Robinson2014}. Uranus and Neptune stand out as notable exceptions, leaving the Ice Giant planetary class unrepresented. A mission to Uranus and/or Neptune would undoubtedly produce data that would be transformed into synthetic exoplanetary observations and used to make predictions and to plan and interpret new observations. 

This series of examples is, however, very specific to transiting exoplanets and there are a number of other ways an atmosphere can be detected such as direct imaging (mentioned in section 2) or phase curve analysis of non-transiting systems (see \cite{Birkby2018book,Biller2018book,sing2018book,Deming2020SpitzerReview}). 
Yet still, all these measurements lack one thing, ground-truth.
With a horde of exoplanet compositions being measured through various remote sensing efforts we lack the ground truth of these worlds that boarder the critical radius regime of the atmospheric transition from that dominated by hydrogen and helium and those with secondary, tertiary, or no atmospheres.

\subsection{Internal structure}
\begin{figure}[!h]
\centering\includegraphics[width=5.2in]{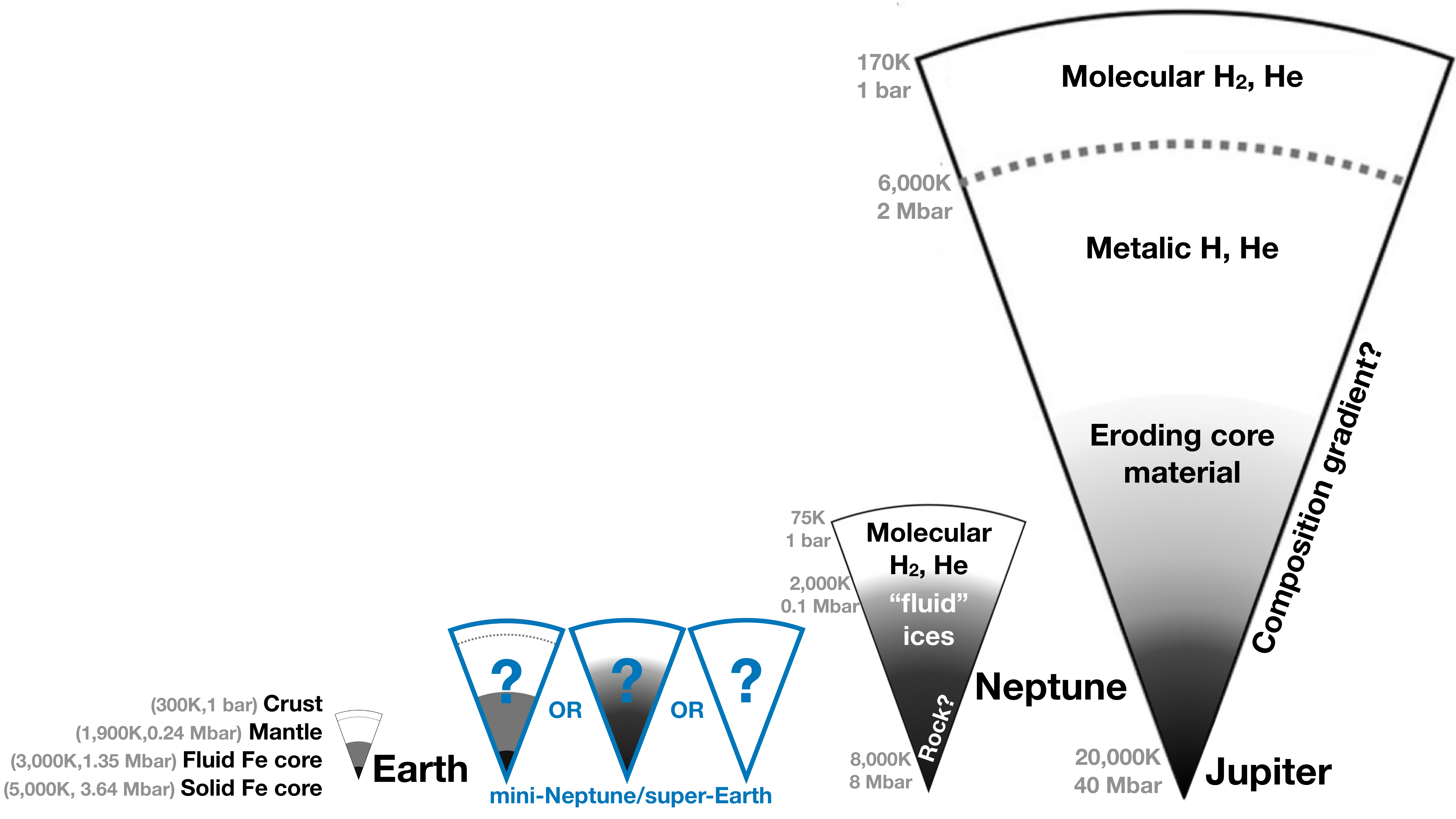}
\caption{Cartoon representation for the interior structure of Jupiter, Neptune, and the Earth (adapted from \cite{spiegel2014}). The cross sections have been scaled to each planets relative radius. The transition between Neptune and Earth has been highlighted where the bulk distribution of exoplanet radii occur with unknown interior structures. Do these sized worlds represent a continuum or is there a step function? What are the outliers and how can we better identify them? Is size/mass the only factor or do other things such as formation location play a role? We also note that the interior structure of Neptune and Uranus is not well known and thus gaining that knowledge will set an upper limit on this transition.}
\label{fig:planet_interior}
\end{figure}
The interior structures of the solar system planets will always be used as templates in exoplanet related applications. For the case of Uranus and Neptune (and their exoplanetary counterparts), the three-layer model---consisting of a rock, ices, and gaseous hydrogen and helium---is a popular interior structure choice (Figure \ref{fig:planet_interior}). Yet, the validity of three-layer structure models has been called into question based on ice-to-rock ratio predictions. Depending on the model, the utilized ice-to-rock ratio ranges from a few to a few dozen, likely an overestimation compared to the stellar value of 2.7 \cite{Nettelmann2013}. 
New analysis of CO and PH$_3$ in Neptune possibly suggests a silicate-rich core, as opposed to an ice- or volatile-rich core \cite{Teanby2019}. This study reconciles D/H and CO measurements for Neptune and removes the requirement for extreme interior O/H enrichment in thermochemical models. It also challenges the very notion that Neptune (and, by extension, Uranus) are ``icy giants.'' These findings and models would certainly benefit from gravitational measurements by an orbital mission to the Ice Giants. The implication of these measurements would also significantly improve the assumptions required to model exoplanets and the statistical results we can obtain from their vast numbers in particular those that lie between the Earth and Neptune in size. 

\begin{figure}[!h]
\centering\includegraphics[width=4.5in]{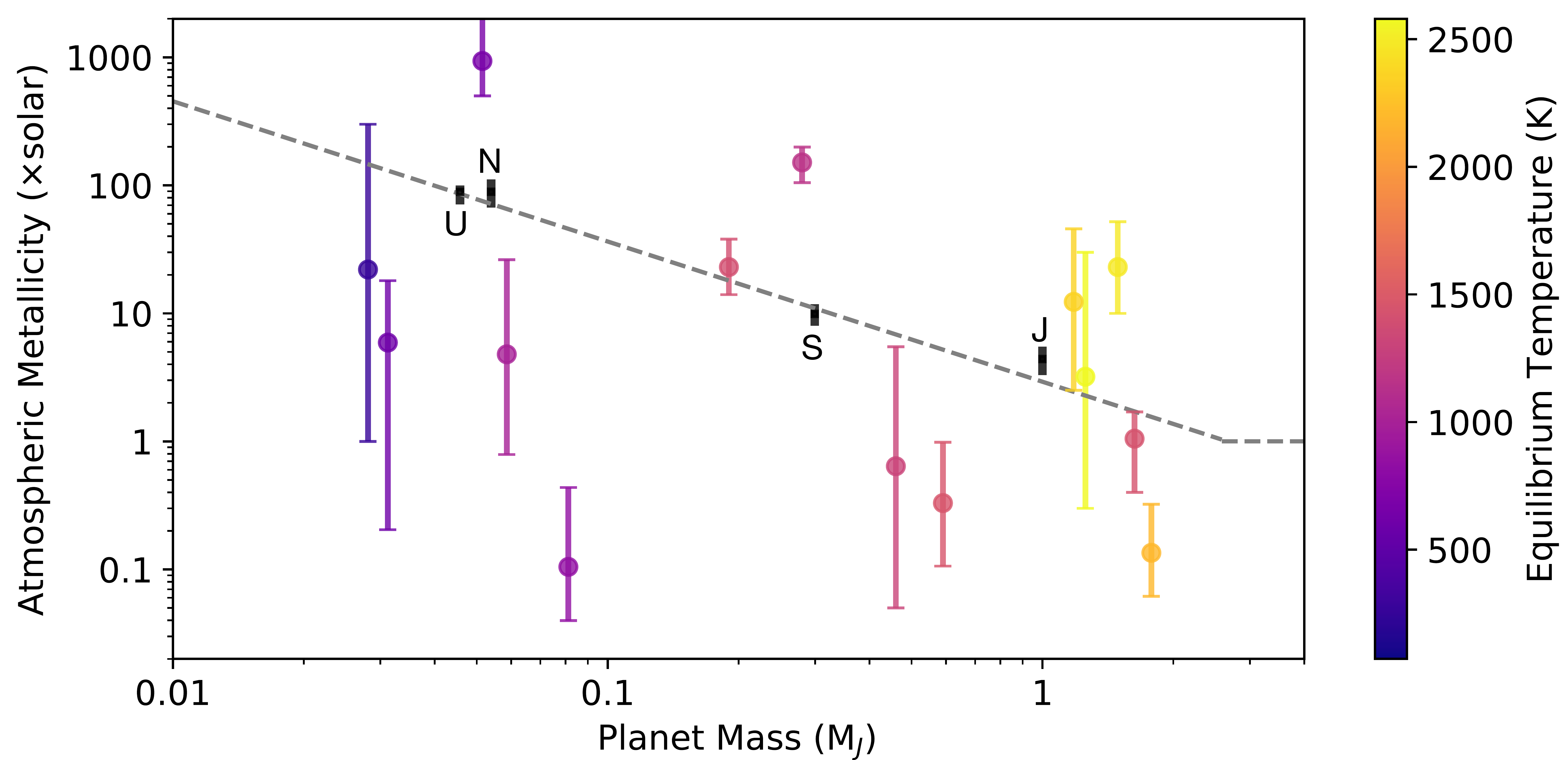}
\caption{Mass-metallicity diagram for the solar system and exoplanets. The solar system measurements \cite{wong2004,fletcher2011,sromovsky2011,karkoschka2011} (black) are based on the CH$_4$ abundance and display a trend increasing the atmospheric metallicity with decreasing planetary mass. The measured exoplanets \cite{mansfield2018,stevenson2017,kreidberg2018,kreidberg2015,Evans2019,Brogi2017,Bruno2020,wakeford2018_w39,spake2020,chachan2019,Wakeford2017Science,morley2017,benneke2019,Benneke2019_K218b} (colored by temperature) do not currently follow this same trend. However, it is important to note that for the exoplanets many are in single planet systems, most if not all migrated to their current position close to their star, and the measurements are predominantly based on H$_2$O absorption or emission features. }
\label{fig:massmetal}
\end{figure}

One study has shown that a planet's bulk metallicity can be used to place an upper limit on the atmospheric metallicity and presents an intriguing possibility of measuring exoplanet atmospheric abundances to infer interior density distributions \cite{throngren2019_atmometal}. However, this requires accurate models of the interior structure of the planet, which do not currently exist for Uranus and Neptune. 
Figure\,\ref{fig:massmetal} shows the planetary mass against atmospheric metallicity trend observed for the solar system whereby the lower the mass the higher the atmospheric metallicity, indicative of formation via core accretion \cite{Pollack1996}. The exoplanet measurements do not currently follow this trend and in many cases defy model predictions for their atmosphere composition and metallicities. The conundrum these data present highlights multiple avenues of improvements that can be made using ground-truth data from Uranus and Neptune. For example: better atmospheric models for ice giant metal-rich atmospheres, interior structure information leading to better models and estimates of the bulk metallicity, information of atmospheric mixing and the impact on measured abundances of carbon and oxygen species, and better understanding of the energy budget of the planet, to name a few. 
There are of course a long list of caveats to the mass-metallicity figure,  namely, directly comparing one system of planets likely all formed together and in relation to each other in their environment, to those of individual planets that have had a significantly different history likely to have migrated through their parent disk to reach their current orbits. The Solar System and exoplanet metallicities each assume that the ratio of heavier elements to the hydrogen content of the atmosphere are equal across all metals. For the Solar System giants the proxy used is carbon measured from CH$_4$ often at specific locations in the atmosphere. Exoplanet studies, on the other hand, predominantly use H$_2$O with oxygen as the proxy, averaging over the entire planet or hemisphere where the comparison is limited by molecule and location. However, the mass-metallicity relation figure can also offer a useful comparison between these two apparently different planetary scenarios. As of now the mass-metallicity relation for exoplanets is limited to a specific subset of exoplanets, but it can be expected that with future measurements more molecules such as CO, CO$_2$, and CH$_4$ will be added to measured abundances for exoplanets, multi-giant-planet systems can be measured, and location specific measurements can be achieved\cite{fortney2005,Powell2019}. On the Solar System side there are improvements to be made as well, while often much harder to obtain different elemental abundances, global-to-local comparisons, and  repeatable measurements can be obtained for our Solar System giant planets adding detail to not only their formation, but their formation in the context of our galactic planet factory.  

The structure of the planet is the central piece in a web of information. For exoplanets the internal structure will impact the observed abundances and material in the upper regions of the atmosphere. The interior structure is also intricately linked to the planets formation and evolution. Measuring, and critically understanding, the interior structure of Uranus and Neptune will be a positive step toward investigation and interpretation of exoplanet observations where gravity measurements will be impossible. 
Without resolving the substantial uncertainty in the fundamental nature of the interiors of the ice giant planets, the prospects for extending interior models to exoplanets are grim. 

\subsection{Formation and evolution}

How the planets in our solar system and beyond formed is one of the big questions in astronomy. Evidence from the solar system its planets, moons, asteroids, comets, and the sun itself have long been used to help understand and answer this question. 
The discovery of exoplanets have now opened up new possibilities for the types of planets, systems, and stars that can form complex and dynamic worlds - with mini-Neptunes and super-Earths dominating the size-distribution. Yet, once again we have little to go on from within our Solar system in regards to the Ice Giants Uranus and Neptune, their systems, rings, moons, that could further facilitate our knowledge of our own existence. 

If Uranus, Neptune, and other intermediate-mass planets form according to the theory core accretion, then they must strike a balance between forming quickly enough to accrete a gaseous envelope but not so quickly that this envelope grows to dominate the mass of the planet. Reproducing the measured masses and solid-to-gas ratios of Uranus and Neptune with standard core accretion models is nontrivial \cite{helled2014}. Small variations in the disk properties, planetesimal properties, and core accretion rate will translate into the formation of of a wide variety of planets. While this outcome is perhaps a useful explanation for the diversity seen amongst exoplanets, it also demonstrates the need for a ground truth for at least one icy giant planet (i.e., Uranus or Neptune). 

The benefit of in-situ measurements to our understanding of giant planet formation and evolution has been witnessed at Jupiter. The Galileo probe made measurements of noble gases that revealed the characteristics of the planetesimals that Jupiter accreted during its formation \cite{Owen1999}. This information can directly inform how models of planet formation treat volatile delivery. Also, the Galileo probe's precise helium measurement provided critical insights as to the thermal evolution of the planet \cite{vonZahn1998}. This finding greatly complements studies of giant exoplanets, especially those observed at different stages in their lifetimes. 

Another critical factor is the formation location of Uranus and Neptune relative to where we see them now. At their present locations, the low density of solids would pose a challenge to formation via core accretion quickly enough to capture some gas before the dissipation of the gas in the disk. Carbon and nitrogen abundance measurements at Jupiter and Saturn paint fairly consistent pictures of giant planet formation via core accretion \cite{wong2004,fletcher2009}. 
However, acquiring similar abundance measurements for Uranus and Neptune is problematic. As in the case of Jupiter and Saturn, the C/H ratio would be measured from the condensable molecule CH$_4$, making any such measurement a lower limit. Similarly, N/H would be measured from NH$_3$, which has also been found to be highly depleted down to at least $\sim$50 bars \cite{Gulkis1978,dePater1989,dePater1991}. This depletion could result from a deep water or superionic ocean \cite{Atreya2020}. With these challenges in mind, it has been found that noble gases will provide the most robust constraints to the formation models of Uranus and Neptune \cite{Atreya2020,Mousis2020}
The outcome would be critical to exoplanets akin to Uranus and Neptune. By their sheer number, there is potential to broadly disentangle current and formation locations based on atmospheric measurements of multiple species. However, having a consistent explanation for Uranus or Neptune to use a waypoint in this effort would be invaluable. 

Lastly, giant impacts have been theorised to greatly change the evolution of a planet \cite{Reinhardt2020}.
The stochastic nature of collisions and giant impacts during formation are potentially a means of increasing the diversity amongst planetary systems \cite{Carter2018,Denman2020}. Giant impacts could indeed explain the bifurcation in the history of Uranus and Neptune \cite{Reinhardt2020}. Refining these theories in Uranus and Neptune would be useful for identifying signs of giant impacts in exoplanet systems.

\section{The next decade for exoplanets}
In the coming decade there are a range of different missions and instruments coming online that will advance our understanding of exoplanets. 
At the start of the 2020's we are in the throes of discovery with the TESS (Transiting Exoplanet Survey Satellite) heading into its extended mission, the ESA CHEOPS satellite starting first light observations, ground-based radial velocity facilities coming online such as Neid on the WIYN Telescope at Kitt Peak National Observatory, the continuation of ground-based transiting exoplanet surveys such as NGTS and SPECULOOS, and through direct imaging with GPI and SPHERE.
In terms of detailed atmospheric characterisation, we have just witnessed \textit{Spitzer}'s final voyage and the loss of a power house of infrared measurements. The Hubble Space Telescope is still dominating molecular and atomic detection's in the atmospheres of close-in exoplanets, and we eagerly await the launch of JWST in the coming years. There are now many ground-based facilities that perform detailed high resolution studies of planetary atmospheres such as CARMENES, CRIRES, and NIRSPEC on Keck II \cite{birkby2018}. The future ELT's will add to this ground-based legacy with high resolution spectroscopy, and advances in direct imagaing studies.
For transiting exoplanets, the ESA ARIEL mission launching in 2028 will add to the \textit{Hubble} and \textit{Webb} legacy of characterization studies by getting spectra of 100's of worlds across a range of sizes, temperatures, and other properties. 
In particular, to highlight studies of Neptune-sized exoplanets, approximately 50\% of Guaranteed Time Observations (GTO) on \textit{Webb} for transiting exoplanets will be spent on sub-Saturnian sized worlds with many more likely being proposed as part of GO time thanks to discoveries from TESS. While, estimates suggest that ESA/ARIEL will photometrically examine over 500 exoplanets between 2\,--\,7 Earth radii\cite{edwards2019ARIEL}.

In the 2030's, the launch of ESA's PLATO telescope will continue the search for ever smaller transiting exoplanets, Gaia will continue to produce accurate astrometry of thousands of stars, and WFIRST could potentially be entering its commissioning phase prior to an all sky survey of exoplanet microlensing events.  
Exoplanets will be well positioned to provide fundamental tests of any planet formation models that derive from solar system observations. The WFIRST survey alone will determine occurrence rates for exoplanets with the mass of Mars or greater at orbital distances from 1--100\,au. While no additional follow-up will be possible for these worlds, their discovery and associated statistical significance cannot be overstated -- there will no longer be a divide in orbital separation between exoplanet occurrence rates and solar system planets. 
In short, we will soon have hundreds of detailed worlds to interpret and understand that will likely spell a revolution in the way we think about our position in the universe.

\section{Conclusion}
The formation of the solar system is a complex puzzle, and at present, Uranus and Neptune are missing pieces. As the number of discovered exoplanets, the amount we learn about their atmospheres, and our knowledge of planetary demographics continues to grow, so will the need to root our exoplanetary discoveries in the foundational knowledge drawn from the solar system. But it is not only these outlined above that rely on this knowledge, the development of our habitable zone also hinges on these fully consistent theories of our solar systems formation and evolution. To ground-truth the Earth and its inhabitants we need to ground-truth the whole system that made it. 
We now know that Uranus and Neptune represent a unique window to exoplanetary systems across the galaxy and thereby play a pivotal role in our quest to answer the question ``{\it How did we get here?}''  
In this way, a mission to the Ice Giants to characterize Uranus and Neptune is part of a larger movement toward placing the entire solar system and its inhabitants in a galactic context.

\enlargethispage{20pt}


\dataccess{Exoplanet archive data can be found here: \href{https://catalogs.mast.stsci.edu/eaot}{Exo.MAST Observability Table}\cite{Mullally2019} 
and \href{https://www.astro.keele.ac.uk/jkt/tepcat/tepcat.html}{TEPCat}\cite{southworth2011}. Models used in Figure 1 can be found here: \href{https://www.cfa.harvard.edu/~lzeng/planetmodels.html\#mrtables}{CfA Harvard planet models}. Data used in Figure 5 can be found here: \href{https://stellarplanet.org/science/mass-metallicity/}{Stellarplanet: mass-metallicity data}}

\aucontribute{Wakeford and Dalba drafted, read, and approved the manuscript. Wakeford composed all the figures based on the referenced data.}

\competing{The authors declare that they have no competing interests.}


\ack{We thank the referees and editor for their helpful comments and insights. Wakeford would like to acknowledge the organisers of the Future exploration of the ice giants meeting at The Royal Society for their invite to present this work and for encouraging outside perspectives on our solar system. Dalba acknowledges support from a National Science Foundation Astronomy \& Astrophysics Postdoctoral Fellowship under award AST-1903811. We also acknowledge the use of Python software tools numpy, scipy, and matplotlib.}


\bibliographystyle{RS.bst}

\end{document}